\documentclass[a4paper,fleqn]{cas-sc}

\usepackage[square,sort,comma,numbers]{natbib}
\usepackage{graphicx}
\usepackage{tabularray}
\def\tsc#1{\csdef{#1}{\textsc{\lowercase{#1}}\xspace}}
\tsc{WGM}
\tsc{QE}
\tsc{EP}
\tsc{PMS}
\tsc{BEC}
\tsc{DE}

\begin{document}
\let\WriteBookmarks\relax
\def\floatpagepagefraction{1}
\def\textpagefraction{.001}
\let\printorcid\relax

\shorttitle{Stock network inference: A framework for market analysis from topology perspective}

\shortauthors{Yijie Teng et~al.}

\title [mode = title]{Stock network inference: A framework for market analysis from topology perspective}   

%
\author[1]{Yijie Teng}[style=chinese]


\credit{Conceptualization of this study, Methodology, Software}

\affiliation[1]{organization={Institute of Fundamental and Frontier Sciences, University of Electronic Science and Technology of China},
    city={Chengdu},
    postcode={610054}, 
    country={P. R. China}}

\affiliation[2]{organization={School of Cyber Science and Technology, University of Science and Technology of China},
    city={Hefei},
    postcode={230026}, 
    country={P. R. China}}

\affiliation[3]{organization={Institute of Dataspace, Hefei Comprehensive National Science Center},
    city={Hefei},
    postcode={230088}, 
    country={P. R. China}}
    
\author[1]{Rongmei Yang}[style=chinese]

\author[3]{Shuqi Xu}[style=chinese]
\cormark[1]
\ead{xushuqi@idata.ah.cn}
\author[1,2,3]{Linyuan L\"u}[style=chinese]

\cormark[1]
\ead{linyuan.lv@uestc.edu.cn}

\credit{Conceptualization of this study, Methodology, Software}

\cortext[cor1]{Corresponding author}

\begin{abstract}
From a complex network perspective, investigating the stock market holds paramount significance as it enables the systematic revelation of topological features inherent in the market. This approach is crucial in exploring market interconnectivity, systemic risks, portfolio management, and structural evolution.
However, prevailing methodologies for constructing networks based on stock data rely on threshold filtering, often needing help to uncover intricate underlying associations among stocks. To address this, we introduce the Stock Network Inference Framework (SNIF), which leverages a self-encoding mechanism. Specifically, the Stock Network Inference Encoder (SNIE) facilitates network construction, while the Movement Prediction Decoder (MPD) enhances movement forecasting. This integrated process culminates in the inference of a stock network, exhibiting remarkable performance across applications such as market structure analysis, stock movement prediction, portfolio construction, and community evolution analysis.
Our approach streamlines the automatic construction of stock networks, liberating the process from threshold dependencies and eliminating the need for additional financial indicators. Incorporating Graph Convolutional Network (GCN) and Long Short-Term Memory (LSTM) models within the SNIF framework, we effectively unearth deep-seated associations among stocks, augmenting the toolset available for comprehensive financial market research. This integration empowers our methodology to automatically construct stock networks without threshold dependencies or reliance on additional economic indicators.
\end{abstract}




\maketitle

\section{Introduction}
In recent years, with the advancement of complex network theory and technology, it has gradually become an effective tool for studying financial markets represented by stock markets. It has been widely used in stock price prediction\cite{jafari2022gcnet,lu2021cnn,ye2021multi}, market risk analysis\cite{zhang2019stability}, and index construction, and has achieved good results.
\par Notably, the network relationships between listed companies are not inherent and necessitate human intervention for their construction. Stock network construction can be categorized into two major classes: explicit networks and implicit networks. Explicit based on real-world contexts that are contingent upon business backgrounds. This class of network construction methods primarily includes building networks based on cross-shareholding\cite{ma2011research}, business interactions\cite{cheng2021modeling}, and economic sectors\cite{gao2021graph} where listed companies operate. 

Implicit network construction methods are based on historical movements of stock prices. This network construction method is based on the efficient market hypothesis (EMH)\cite{fama1970efficient}, a cornerstone of modern financial theory. The EMH asserts that asset prices in financial markets reflect all available information. Although the fractal market hypothesis (FMH)\cite{peters1994fractal}, which is based on nonlinear dynamical systems, can better explain modern financial market phenomena, the EMH still has practical reference and guidance due to the mathematical difficulties of modeling the FMH. By extending market price information to interconnections, the relationships between listed companies can also be established through the movement of stock prices.

The most widely used network construction method based on historical stock price data is the logarithmic return correlation coefficient method, which determines the links between stocks\cite{heiberger2014stock,chi2010network,huang2009network}. In recent years, researchers have developed more innovative approaches to constructing stock networks to investigate financial markets from various perspectives. For instance, Ya-Chun Gao et al.\cite{gao2015influence} completed directed links between stocks through time-dependent cross-correlation and analyzed the impact relationships among economic sectors. Wei-Qiang Huang et al.\ note {huang2016financial} used the dynamic conditional correlations (DCC) method to construct links. Then, it filtered the resulting graph using the minimum spanning tree (MST) to analyze the risk contribution of listed companies to the market. Weiping Zhang et al.\cite{zhang2019stability} used multiscale detrended fluctuation analysis (MSDFA) and value-at-risk (VaR) to construct a multiscale curve network and a risk network and analyzed the stability of financial markets. Yuta Arai et al.\cite{arai2015dynamic} employed complex principal component analysis (CPCA) to separate the market volatility central component and constructed a network to analyze the communities within the stock network.

The meaningful stock network structure based on similarity indexes derived from stock price movements are usually fully connected graphs and, therefore, require filtering to reduce the density generally. Traditional studies usually use threshold filtering, i.e., filtering out edges with similarity or other metrics less than a given threshold\cite{arai2015dynamic}. Other graph filtering methods have also been proposed, such as minimum spanning tree (MST) filtering\cite{wang2012similarity,djauhari2013minimal}, and planar maximally filtered graph (PMFG)\cite{wang2017multiscale}. However, these filtering algorithms either depend on the artificial selection of a threshold or lose the higher-order structure(networks generated by MST have no circles, and networks developed by PMFG have only triangles) of the network, making it challenging to construct a natural and informative stock network. In recent years, methods for inferring the explicit interaction structure of dynamic systems using deep learning and graph neural networks have become popular\cite{kipf2018neural}. It allows us to discover the interactions of objects in the real world with high precision in an unsupervised manner and accurately predict the dynamics of future time steps. However, this effective method has yet to be used for relationship mining in financial markets.

In this work, we propose SNIF, an innovative use of deep learning models to build stock networks. We use a CNN-based Encoder on multidimensional stock price data to compute stock node embeddings and reason about potential relationships between stocks. We use GCN and LSTM for stock movement prediction as Decoder modules of SNIF to help better learn the relationship between stocks. Finally, we apply stock movement prediction, portfolio selection, network topology, and community evolution analysis. The experimental results show that the network inferred by SNIF can fully automatically imply the potential relationship between stocks while retaining a rich network topology. This can help us better understand the complex financial dynamic system.



\section{Method}
\subsection{Stock Network Inference Framework}
\begin{figure}[htbp]
\label{framework}
\centering
\includegraphics[width= \linewidth]{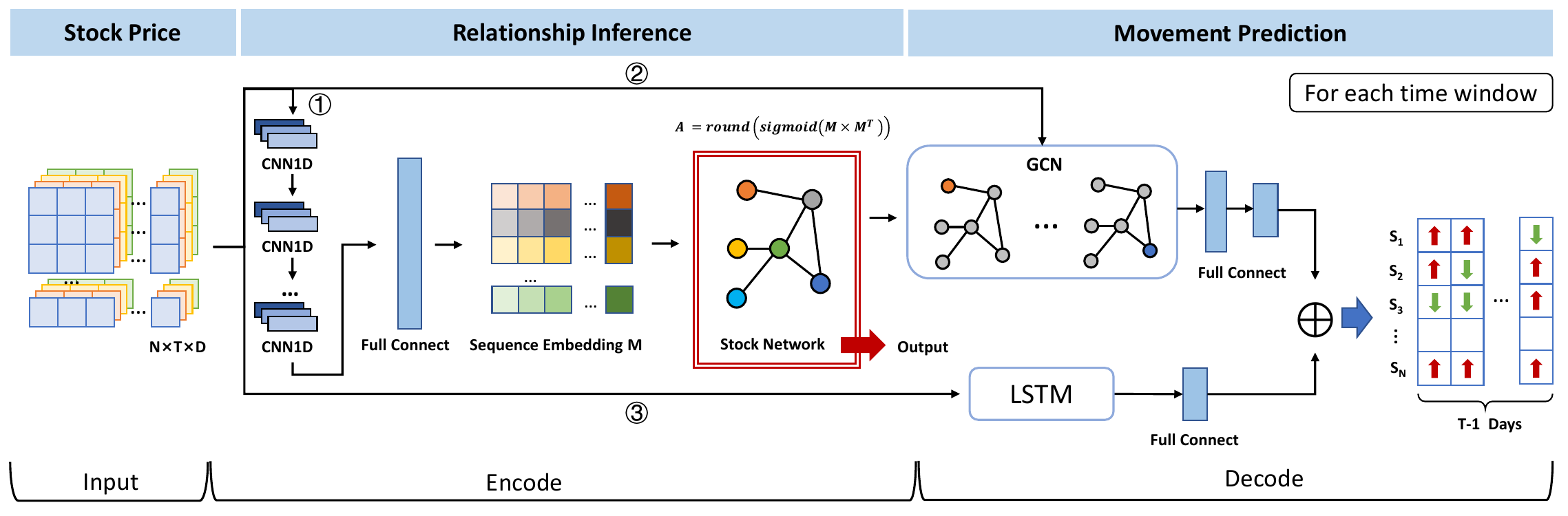}
\caption{{\bf Stock Network Inference Framework.} The SNIF is primarily composed of two integral modules: SRIE and MPD. The SRIE employs a composition of several 1D CNN layers and a fully connected layer to extract essential price movement features from stock nodes. This process facilitates the extraction of inter-node relationships and the construction of the network. In parallel, the MPD integrates elements of GCN, LSTM, and fully connected layers. This configuration harnesses the message passing of GCN to enhance node attributes, simultaneously considering the temporal dynamics inherent in price sequences to predict the movements of stock prices for the subsequent trading day. In terms of input processing, there are three price data streams: (1) By passing through the SRIE, the data yields a node sequence embedding matrix $M$, which subsequently serves as the foundation for generating the stock network adjacency matrix $A$ through $M \times M^T$. (2) The initial price data undergoes inter-node message propagation via GCN in conjunction with the network derived from step (1). This process effectively captures intrinsic features, contributing to the first facet (network-structure-based) of movement prediction through two subsequent fully connected layers. (3) The initial price data undergoes temporal feature extraction through LSTM and fully connected layers, culminating in the second facet (time-series-based) of movement prediction. The result of amalgamating (2) and (3) can predict the movement of the stock sequences. Upon completion of training on the designated training dataset and the fine-tuning of hyperparameters on the validation set, the resultant adjacency matrix A derived from step (1) serves as a foundational representation for the stock inference network. 
}
\end{figure}

The Stock Network Inference Framework(SNIF) employs a structure based on autoencoder\cite{bank2020autoencoders}, artificial neural networks designed for unsupervised learning tasks. The overall framework is schematically depicted in Figure\ref{framework}. An autoencoder consists of an encoder component responsible for mapping input data into a lower-dimensional latent space representation, followed by a decoder component that reconstructs the original input data from this reduced representation. This architecture is commonly utilized for dimensionality reduction, denoising, and anomaly detection tasks. In the context of SNIF, applying autoencoders contributes to the efficient extraction and representation of stock market network features. SNIF comprises two components: the Stock Relationship Inference Encoder and the Movement Prediction Decoder. The former learns the feature representations of individual stocks from the sequences of stock price movements and establishes connections among them. The latter leverages the established relationships and the integration of GCN and LSTM to predict stock price movements. The matrix $M$ is then transformed into an adjacency matrix A representing the stock network. The improved accuracy of movement predictions indicates the effectiveness of the network constructed by the Stock Relationship Inference Encoder.

{\bf Stock Relationship Inference Encode.}
The Stock Relationship Inference Encoder (SRIE) applies several layers of 1D CNN (specifically, two layers) and neural network (NN) feature extraction to the preprocessed daily stock price sequences to capture the interdependencies among the stocks. In applying SRIE, the entire training dataset is utilized as the primary input for relation encoding. This process results in a feature representation matrix $M$, which serves as an embedding for the stocks. It is important to note that the resulting adjacency matrix $A$ is symmetric since the correlation between stocks is considered bidirectional.

More formally, let us define the input data as $S=\left (  s_{itd}\right ) _{N\times T \times D} $, where $s_{itd}$ denotes the price of the stock $i$ on the day $t$ in the $d$-dimension price. Here, $N$ represents the number of stocks, $T$ represents the count of trading days, and $D$ represents the number of price dimensions.
For a given input $S$, the adjacency matrix $A$ is computed through the following process.
\begin{equation}
    \begin{aligned}
        M &= f_{emb,\phi}(S),\\
        A &= \sigma (M\times M^T).
    \end{aligned}
\end{equation}
In this context, the functions $f_{emb,\phi}$ are dependent on the model parameter $\phi$, facilitating the inference of features of input $S$. The resulting matrix $A$ represents the adjacency matrix of the stock network, derived from encoding processes.

{\bf Movement Prediction Decoder.} The Movement Prediction Decoder (MPD) leverages the stock relationship network constructed based on SRIE. It incorporates LSTM to capture the temporal dynamics of individual stocks and integrates GCN to extract relational features from the stock network. Previous studies have indicated that considering the stock network structure can improve prediction accuracy\cite{hou2021st,sawhney2020spatiotemporal}. The output layer of the MPD is utilized to predict the upward or downward movement of the stock for the next trading day, treating it as a binary classification problem. The performance of movement prediction reflects the rationality and effectiveness of the stock network construction process through SRIE.

Assuming we have the input data $S_{1:N,1:T,1:D}$. We aim to predict $Tr_{1:N, T+1, C}$, which corresponds to the price movement (rise or fall) of $N$ stocks on the day $(T+1)$ relative to the before $T$ trading day, MPD prediction is accomplished through the following procedure.
\begin{equation}
     Tr =softmax(P_1 + P_2) =softmax(f_{gra,\phi_1}(S, A) + f_{lstm,\phi_2}(S)),
\end{equation}
where $Tr$ refers to the predicted results, and$f_{gra,\phi_1}(S, A) $ constitutes the module that derives inferences utilizing GCN as the primary component under the parameter configuration $\phi_1$. Similarly, $f_{lstm,\phi_2}(S)$ represents the module that produces inferences employing LSTM as the primary component under the parameter $\phi_2$. These two segments independently generate partial predictive outcomes, denoted as $P_1$ and $P_2$ (of size $N \times 2$). These predictive outcomes are then combined and normalized using the softmax function to yield normalized predictive probabilities.

\subsection{Graph Convolutional Network}
Graph Convolutional Networks (GCN) are a type of neural network that can operate on graph-structured data. They have become popular in recent years for their ability to learn representations of nodes in a graph, which can then be used for various downstream tasks such as node classification, link prediction, and community detection.
The convolution operation in GCN is defined as:
\begin{equation}
H^{(l+1)} = \sigma(\hat{D}^{-\frac{1}{2}}\hat{A}\hat{D}^{-\frac{1}{2}}H^{(l)}W^{(l)}),
\end{equation}
where $H^{(l)}$ is the hidden state at layer $l$, $\hat{A}$ is the normalized adjacency matrix, $\hat{D}$ is the degree matrix, $W^{(l)}$ is the weight matrix at layer $l$, and $\sigma$ is the activation function. This operation aggregates the information from a node's neighbors and updates the node's representation at each layer.
\subsection{Long Short-Term Memory}
Long Short-Term Memory (LSTM) is a type of recurrent neural network (RNN) architecture that has gained significant attention and popularity in deep learning. It was specifically designed to overcome the limitation of standard RNNs in capturing and modeling long-range dependencies in sequential data.

The critical innovation of LSTM lies in its ability to maintain and control information flow over time through a gated memory structure. This memory structure consists of memory cells and three types of gates: the input, forget, and output. These gates regulate the flow of information into, out of, and within the memory cells.

The LSTM's memory cell retains information for an extended period, selectively updating and forgetting information as needed. This characteristic enables LSTM to capture long-term dependencies and address the vanishing or exploding gradient problem that can hinder the training of traditional RNNs.

Mathematically, the LSTM equations can be described as follows:

\begin{equation}
    \begin{aligned}
        i_t &= \sigma(W_{xi}x_t + W_{hi}h_{t-1} + W_{ci}c_{t-1} + b_i), \\
        f_t &= \sigma(W_{xf}x_t + W_{hf}h_{t-1} + W_{cf}c_{t-1} + b_f), \\
        c_t &= f_tc_{t-1} + i_t\tanh(W_{xc}x_t + W_{hc}h_{t-1} + b_c), \\
        o_t &= \sigma(W_{xo}x_t + W_{ho}h_{t-1} + W_{co}c_t + b_o), \\
        h_t &= o_t\tanh(c_t).
    \end{aligned}
\end{equation}
Here, $x_t$ represents the input at time step $t$, $h_t$ denotes the hidden state at time step $t$, $c_t$ means the cell state at time step $t$ and $\sigma$ represents the sigmoid function. The weight matrices $W$ and bias vectors $b$ are learnable parameters of the LSTM network.

\section{Result}
\subsection{Data Description}
\begin{figure}[htbp]
\centering
\includegraphics[width=1 \linewidth]{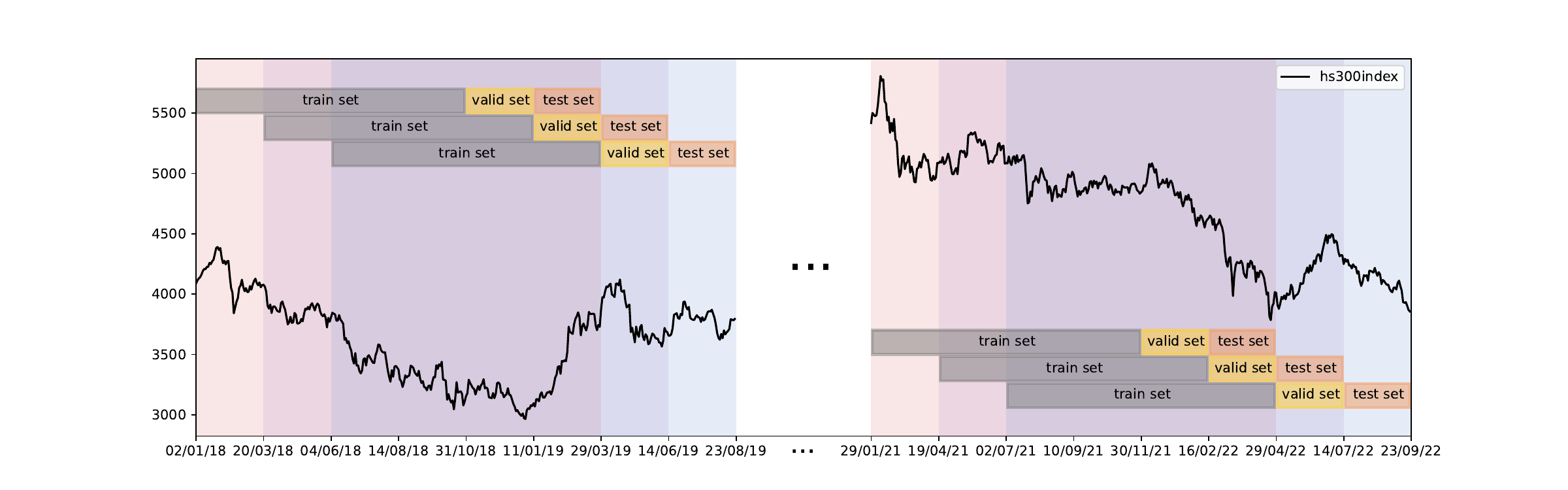}
\caption{{\bf Time window.} We applied a sliding window approach with a step size of 50 trading days to select a continuous training set, validation set, and test set from the 171 stocks identified in the CSI 300 Index. The training set consisted of 200 trading days, the validation set consisted of 50 trading days, and the test set consisted of 50 trading days. Using overlapping sliding windows, we obtained 18-time windows from the 1150 trading days dataset. For each of these 18 time windows, we utilized the Stock Relationship Inference Encoder (SRIE) to infer the stock networks at different periods.}
\label{fig:time_span}
\end{figure}

{\bf Data collection.}
The CSI300 index comprises 300 stocks listed on the Shanghai Stock Exchange and the Shenzhen Stock Exchange. Widely employed in studying China's financial markets, it was initially introduced on April 4, 2005, as a comprehensive indicator of the Chinese stock market's overall performance \cite{xia2018comparison}. For our research, we select the constituent stocks of the CSI300 index for 2022 and ensure that the stocks chosen did not have consecutive trading suspensions exceeding a single day. In addition, considering factors such as stock market capitalization and number of trading days, we ultimately selected 171 stocks spanning from January 1, 2018, to September 23, 2022, encompassing 1,150 trading days. Furthermore, the data is partitioned into overlapping time windows 300 (see Figure. \ref{fig:time_span}). Each time window is divided into three subsets: the first 200 trading days serve as the training set, the subsequent 50 trading days (201-250) constitute the validation set, and the final 50 trading days (251-300) form the test set.

{\bf Data preprocessing.} For trading days when stock trading was suspended due to trading suspensions, missing data points are filled using linear interpolation. Additionally, to account for corporate actions such as stock splits, dividends, and bonus issues, a unified treatment is applied using the following formula for forward adjustment:
\begin{equation}
P_{\text{ex}} = \frac{P_{\text{record}} - D}{1 + B}
\end{equation}

Here, $P_{\text{ex}}$ corresponds to the ex-rights ex-dividend price, $P_{\text{record}}$ stands for the price information on the record date, $D$ represents the dividend per share, and $B$ signifies the number of bonus shares per share.

Within each time window, the prices are normalized using the Min-Max Normalization, represented by the following formula:
\begin{equation}
P_{i}^{new} = \frac{P_{i} - P_{min}}{P_{max} - P_{min}}.
\end{equation}
In this formula, $P_{i}$ represents the original price of the stock $i$, $P_{min}$ corresponds to the minimum price within the time window, and $P_{max}$ denotes the maximum price within the time window. By applying this normalization formula, the prices are scaled to 0 to 1, facilitating meaningful comparisons and analysis.

\subsection{Stock Movement Prediction}
As detailed in Section 2.1, SNIF employs the prediction of stock trends to infer relationships among stocks, thereby constructing a network. Predicting stock trends constitutes a common task within stock forecasting, aimed at anticipating the upward or downward movement of stocks for the following day, which holds significant implications within financial markets. In prior research endeavors\cite{zhu2020stock,althelaya2018stock}, LSTM, RNN, and GRU have been frequently employed for stock price prediction. Consequently, we conducted comparative analyses of the three classical deep learning models, namely LSTM, RNN, and GRU, along with their corresponding performance under the SNIF framework. Our evaluation used comprehensive performance metrics such as AUC (Area Under the Curve), ACC (Accuracy), and Precision.

The Area Under the Curve (AUC) is a widely used metric for evaluating the performance of binary classification models. It comprehensively measures the model's ability to discriminate between positive and negative instances across different classification thresholds. A higher AUC value indicates better model performance. The formula is as follows:

\begin{equation}
AUC = \frac{\sum(p_i,n_j)_{p_i>n_j}}{P \cdot N},
\end{equation}
where $P$ denotes the number of positive samples, $N$ signifies the number of negative samples, $p_i$ represents the predictive score of positive samples, and $p_j$ corresponds to the predictive score of negative samples.

Accuracy is a fundamental metric that measures the proportion of correctly predicted instances to the total number of cases in the dataset. It provides a general overview of the model's overall performance. Precision is a metric that focuses on the proportion of true positive predictions among all positive predictions made by the model. It is beneficial when the cost of false positives is high. The formula of ACC and Precision are as follows:

\begin{equation}
ACC = \frac{TP + TN}{TP + TN + FP + FN},
\end{equation}

\begin{equation}
Precision = \frac{TP}{TP + FP},
\end{equation}
where $TP$ is the number of true positives, $TN$ is the number of true negatives, $FP$ is the number of false positives, and $FN$ is the number of false negatives.

\begin{figure}[htbp]
\centering
\includegraphics[width= \linewidth]{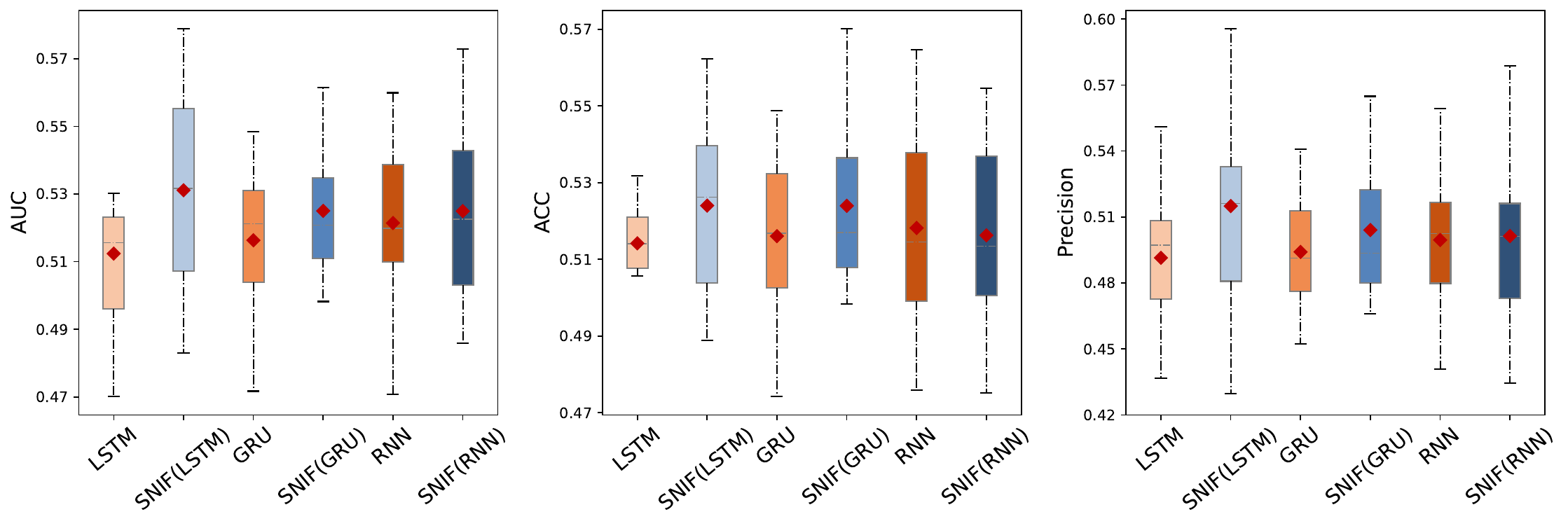}
\caption{{\bf The performance of temporal stock network movement prediction decoder.} We compare the ACC, AUC, and Precision indicators of movement prediction between the MPD model and the baseline LSTM model across 18 data stages. }
\label{fig:trend_predict}
\end{figure}

It is worth noting that the primary objective of SNIF does not lie in forecasting stock trends (instead, it serves as an auxiliary outcome derived from the construction of the stock network). However, SNIF has demonstrated commendable performance in this regard as well. Table \ref{tab:trend_predict6} presents the averaged outcomes of stock trend prediction across 18 phases, encompassing AUC, ACC, and Precision. The analytical results indicate that, following the incorporation of SNIF, except for a marginal difference in ACC between SNIF(RNN) and RNN, all other groups exhibited comprehensive enhancements. SNI(LSTM) showed the most promising outcomes, with ACC, AUC, and Precision scores of $0.5240$, $0.5311$, and $0.5150$, respectively. Furthermore, we have also depicted the outcomes of the 18-time windows using box plots to vividly manifest the overall superiority of the SNIF framework, as illustrated in Fig. \ref{fig:trend_predict}.

We must reiterate that our primary focus lies in network construction rather than individual stock trend prediction. Consequently, we compared LSTM, RNN, and GRU as the respective benchmark models. To ensure comparability, we maintained uniformity across LSTM, RNN, GRU, and the corresponding SNIF framework models by employing identical iterations, early stopping strategies, optimizers, learning rates, and other hyperparameters within the pertinent sections.

\begin{table}[htbp]
\centering
\renewcommand\arraystretch{1.5}
\caption{{\bf The performance of temporal stock network movement prediction decoder.} The best results are shown in bold.}
\label{tab:trend_predict6}
\begin{tabular}{cccc} 
\hline
Model     & AUC              & ACC              & Precision         \\ 
\hline
LSTM &  0.5124          & 0.5142           & 0.4915            \\
SNIF(LSTM) & \textbf{0.5311 }  & \textbf{0.5240 } & \textbf{0.5150 }  \\ 
\hline
GRU  & 0.5163           & 0.5160           & 0.4941            \\
SNIF(GRU) & \textbf{0.5250}  & \textbf{0.5240}  & \textbf{0.5041}   \\ 
\hline
RNN  &  0.5214         & \textbf{0.5182 }           & 0.4995            \\
SNIF(RNN) & \textbf{0.5249} & 0.5163   & \textbf{0.5014}   \\
\hline
\end{tabular}
\end{table}

\subsection{Network Topology Structure}
We considered the correlation between specific topological characteristics of the stock inference network and the overall movement dispersion. These topological characteristics encompass the average degree, clustering coefficient, average path length, and assortativity\cite{boccaletti2006complex}. The general movement dispersion is computed within the trading day intervals corresponding to the network structure. Worth noting is that for each stock, we construct training and validation subsets as the scope for calculating the movement dispersion, given their involvement in the construction of the stock inference network. We employed Dynamic Time Warping(DTW)\cite{muller2007dynamic} to assess the disparity between individual stock's cumulative returns and the overall cumulative returns and then calculated the average to measure the stock returns separation level within a specific time interval. Precisely, the general movement dispersion is calculated as follows: 

Step 1: Calculation of Weighted Index
Let us consider a set of n stocks denoted by $S_1$, $S_2$, ..., $S_n$, with their corresponding market capitalizations represented by $M_1$, $M_2$, ..., $M_n$. The weighted index, denoted as $I_t$, for a particular day $t$, is computed using the formula:
\begin{equation}
I_t = \frac{\sum_{i=1}^{n}M_i\cdot P_{i,t}}{D_t},
\end{equation}
where $M_i$ is the market capitalization of stock $S_i$, $P_{i,t}$ is the price of stock $S_i$ on day $t$, and $D_t$ is a normalization factor.

Step 2: Calculation of Returns Series
Having obtained the weighted index values for each day, the returns series, denoted as $R_t$, for the day $t$ is computed as follows:

\begin{equation}
R_t = \frac{I_t - I_1}{I_1},
\end{equation}
where $I_t$ is the index value on day t, and $I_1$ is the index value on the first day of the time series.

Step 3: Calculation of DTW Values between Individual Stock Returns and $R_t$

The Dynamic Time Warping (DTW) value for each stock is calculated to quantify the dissimilarity between its individual daily returns and the previously computed returns series, denoted as $R_t$. Let $R_i = [r_i(1), r_i(2), ..., r_i(m)]$ be the returns series of the $i-th$ stock, and $R_t = [r_t(1), r_t(2), ..., r_t(n)]$ be the returns series of the overall weighted index, $R_t$.
The DTW distance between $R_i$ and $R_t$, denoted as $DTW(R_i, R_t)$, is calculated as follows:

\begin{equation}
DTW\left(R_i, R_t\right)=\sqrt{\sum_{k=1}^n \sum_{j=1}^m\left(r_i(j)-r_t(k)\right)^2 \cdot \omega(j, k)},
\end{equation}
where $\omega(j, k)$ is the warping path weight, which is determined by the dynamic programming approach of DTW, the warping path weight represents the alignment cost between the $j-th$ data point of $R_i$ and the $k-th$ data point of $R_t$.

Step 4: Calculation of Average DTW Values for Interval Movement Dispersion

In the fourth step, we compute the average of the DTW values obtained in Step 3 for all stocks within the given time interval. Denoting the number of stocks as n, the average DTW value, represented as $DTW_{avg}$, is calculated as follows:

\begin{equation}
DTW_{avg}=\frac{1}{n} \sum_{i=1}^n D T W\left(R_i, R_t\right).
\end{equation}

The resulting $DTW_{avg}$ measures the movement dispersion or dissimilarity between the individual stock returns and the overall market index returns within the specific time interval. A higher $DTW_{avg}$ value indicates a greater degree of divergence or separation between the stock returns and the market index. In contrast, a lower value suggests a higher level of alignment or similarity.

\begin{figure}[htbp]
\centering
\includegraphics[width=0.5\linewidth]{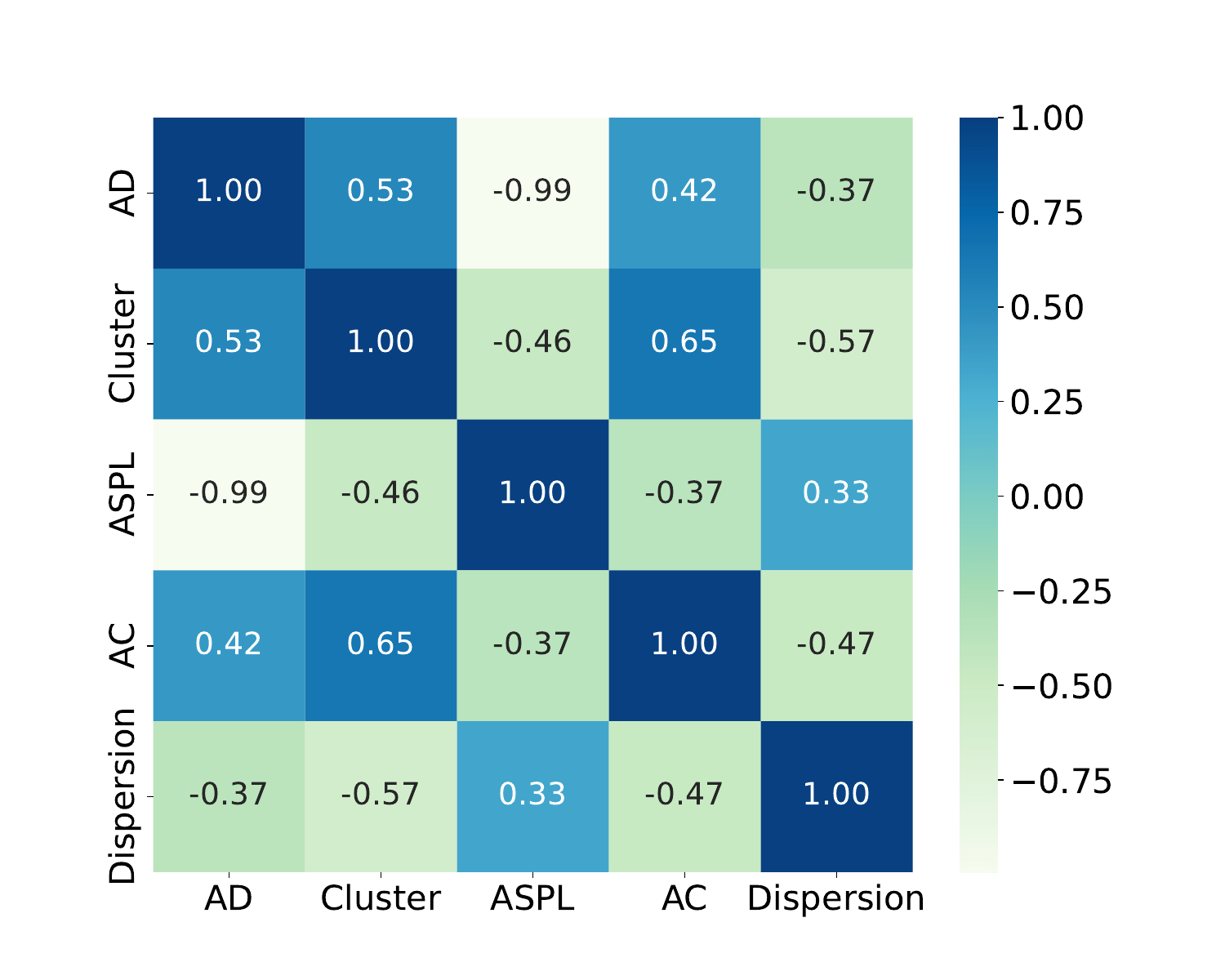}
\caption{{\bf The correlation between topological metrics and movement dispersion.} As depicted in the figure, the movement dispersion of stocks exhibits a positive correlation with the average degree and clustering coefficient while demonstrating a negative correlation with the average path length. This observation aligns with intuition, as the constructed network's nodes tend to disperse rather than concentrate during movement dispersion in stocks. Additionally, the dispersion degree shows a negative correlation with degree assortativity, indicating that the relationships between stock nodes tend to be disassortative in times of movement dispersion.}
\label{fig:corr}
\end{figure}
We analyze the dispersion $DTW_{avg}$ computed for each stage alongside the corresponding network topological metrics, namely average degree, clustering coefficient, average path length, and degree assortativity, as mentioned earlier. This analysis aims to establish the relationship between the overall movement dispersion of stocks and the characteristics of the stock inference network. The results are presented in Figure \ref{fig:corr}.

\subsection{Portfolios Analysis}

\begin{figure}[htbp]
\centering
\includegraphics[width=1 \linewidth]{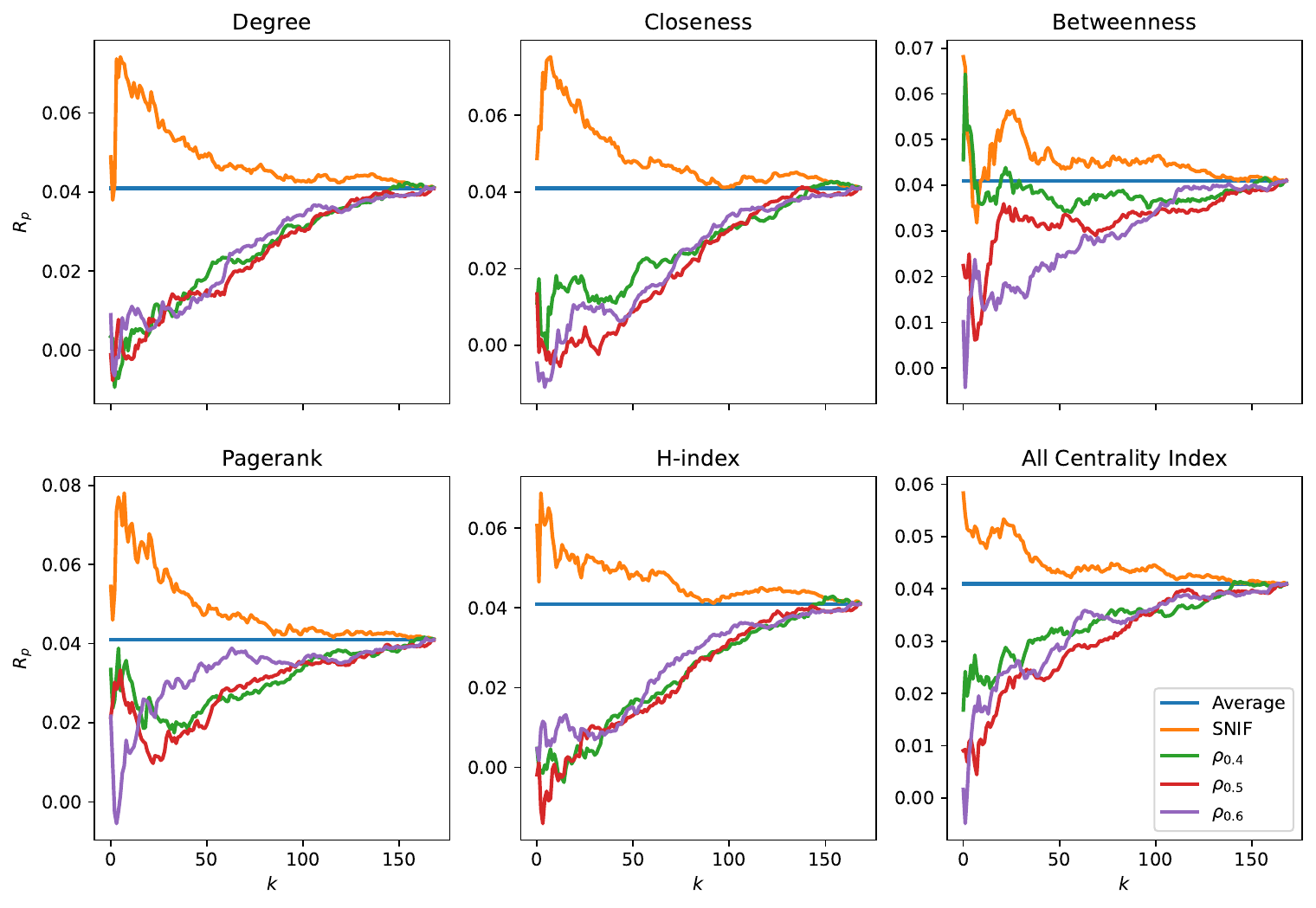}
\caption{{\bf The rate of returns of different portfolios.} The $k$ on the $x$-axis represents the number of stocks selected for each centrality index, and the $y$-axis represents the rate of return. We considered portfolios with at least two stocks for analysis. Therefore, for the single centrality index, the range of $k$ is from 2 to $N$, while for all centrality index, the range of $k$ starts from 1.}
\label{fig:portfolios}
\end{figure}

We also explored the application of the stock inference network in a typical financial market context, namely, portfolio construction. Portfolio construction involves selecting a subset of stocks from the stock pool to form a portfolio to achieve excess returns while balancing market risk\cite{constantinides1995portfolio}. Differing from traditional portfolio construction methods\cite{ma2008return,bertsimas1999portfolio,korzeniewski2018efficient,mishra2019efficient}, we leverage the topological structure of the stock inference network to select indicators and identify essential nodes within the stock network as constituents of the stock investment portfolio. We investigated the average returns of portfolios over multiple periods and varying stock scales within fixed intervals.
To simplify the problem, we only backtested the average returns under different network construction methods, considering the following two assumptions:

Assumption 1: The capital is appropriately allocated and can be evenly distributed among each stock in the investment portfolio.
Assumption 2: The buying and selling of these individual stocks do not significantly impact their respective movements, assuming the capital flows are not substantial.

Specifically, for each stock inference network construction time interval corresponding to the data interval of the inference network, we utilized the test interval as the return backtesting period for the investment portfolio. We examined the average returns over 18 segments from the beginning of the test interval (holding the investment portfolio stocks equally) to the end of the test interval (50 days later). The average return of the investment portfolio at time $t$, $R_p$, can be calculated using the following formula:
\begin{equation}
R_{p}=\frac{1}{N_w} \cdot\sum_{T_d}^{T_d+N_w} \sum_{i=1}^{n}\frac{p_i(T_d+C) - p_i(T_d)}{p_i(T_d)},th
\end{equation}
where $N_w$ represents the number of backtesting intervals, $T_d$ denotes the starting trading day of the $d$th backtesting interval, 
$C$ represents the length of the backtesting interval, and $p_i(T)$ denotes the price of the stock $i$ in the investment portfolio on the trading day $T$.

Regarding the selection of constituent stocks for the investment portfolio, we considered various centrality measures\cite{newman2018networks,lu2016h}, namely degree centrality, closeness centrality, betweenness centrality, PageRank, and H-index, either individually or in combination. The top $k$ stocks based on each centrality measure were selected from the stock inference network to form the union of the constituents of the investment portfolio. Furthermore, as a point of comparison, we also employ the classical approach to construct the stock network, using the correlation coefficient of logarithmic returns to represent the similarity between stocks. We selected three threshold values to filter the networks, namely $t=0.4, 0.5$, and $0.6$, to generate three stock return correlation networks. Subsequently, we performed the aforementioned node centrality calculations to select investment portfolios based on these correlation networks. These portfolios were then compared with the investment portfolio derived from our inference network regarding their respective returns, resulting in the findings presented in Figure \ref{fig:portfolios}.

From Figure \ref{fig:portfolios}, it can be observed that the portfolio returns of the SNIF outperformed the average (using all stocks, represented by the blue line) and the stock return correlation networks in most cases of $k$ when considering different centrality metrics. When the value of $k$ is small, SNIF and the stock return correlation network($p_{0.4}$) have similar results in betweenness centrality, and SNIF results are better when $k$ is increased. When considering single metrics and lower values of $k (k<5)$, portfolio returns may fluctuate significantly. This is due to the smaller portfolio size, where individual stock performance has a more pronounced impact. This issue is overcome when the portfolio is selected using a combination of all centrality metrics, resulting in more stable returns (see Figure \ref{fig:portfolios}, lower right subgraph).

\subsection{Community Evolution Analysis}
In this section, we analyzed the generated 18 networks to identify and study their communities, as well as the evolution of these communities over time. Additionally, we investigated the industry distribution and flow patterns within each community. We employed the Louvain algorithm for community detection, a widely used and efficient method for detecting communities in complex networks. Briefly, the Louvain algorithm\cite{louvain} is a modularity optimization-based method that aims to maximize the modularity of the network. It iteratively optimizes the modularity by moving nodes between communities until a locally optimal modularity value is reached. The algorithm efficiently identifies community structures in large-scale networks and has been widely applied in various fields, including social network analysis and biological network analysis.

\begin{figure}[htbp]
\centering
\includegraphics[width=0.6 \linewidth]{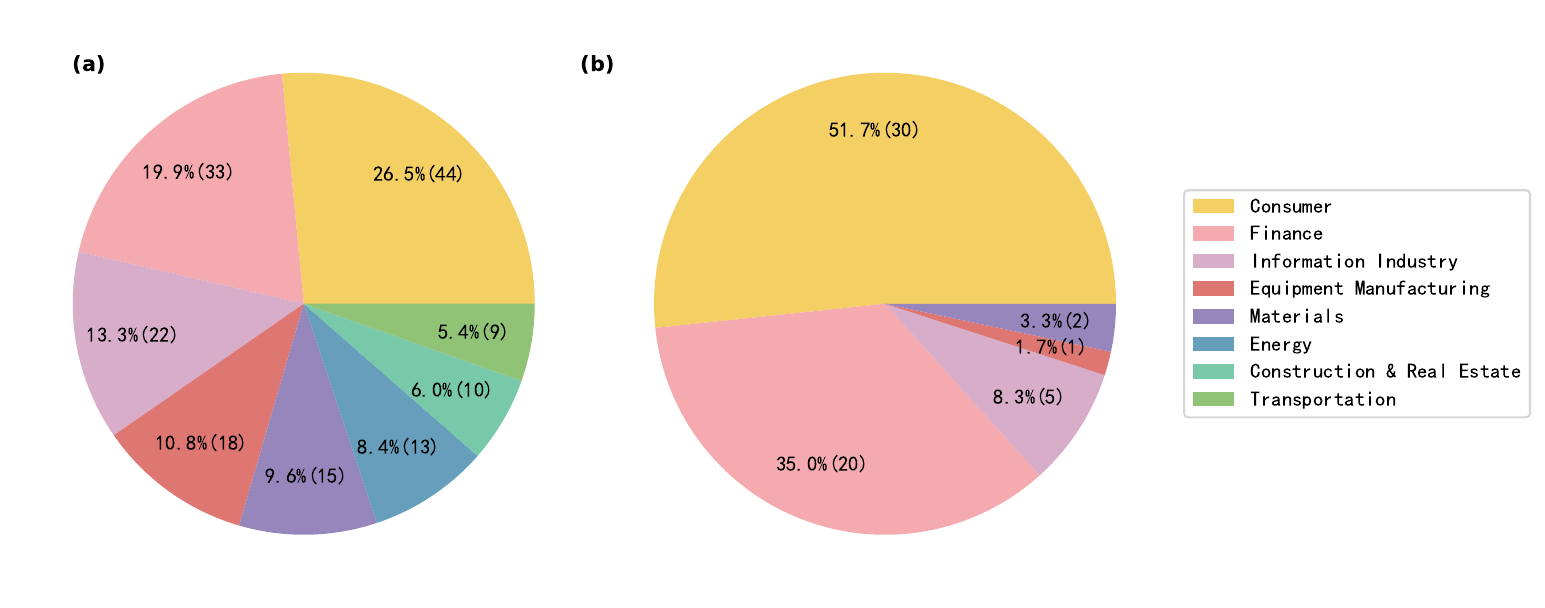}
\caption{{\bf The distribution of the industry.} {\bf (a)Industry distribution of the covered public companies in this study. } {\bf (b) The distribution where industries dominate the majority of communities.} If stocks from a particular industry hold the highest count within a community, that community is considered to be led by the dominant industry.}
\label{fig:industry_distribution}
\end{figure}

Regarding the industry classification, we utilized the same scheme as the data source, specifically the Tongdaxin first-level industry classification. Based on the correspondence between the 171 screened stocks and the industry classification, we excluded sectors with insufficient stocks and merged similar sectors. As a result, we identified eight primary sectors, which include Finance, Construction \& Real Estate, Information Technology, Equipment Manufacturing, Consumer, Materials, Transportation, and Energy (Stocks and their corresponding sectors are shown in Table. \ref{tab:stock_ticker}). The number of nodes contained within each department is relatively balanced, as illustrated in Figure \ref{fig:industry_distribution} {\bf (a)}. However, the predominant industries within each association (those with the most nodes) vary. Across all temporal communities, the Consumer and Finance sectors hold the majority.

\begin{figure}[htbp]
\centering
\includegraphics[width=1 \linewidth]{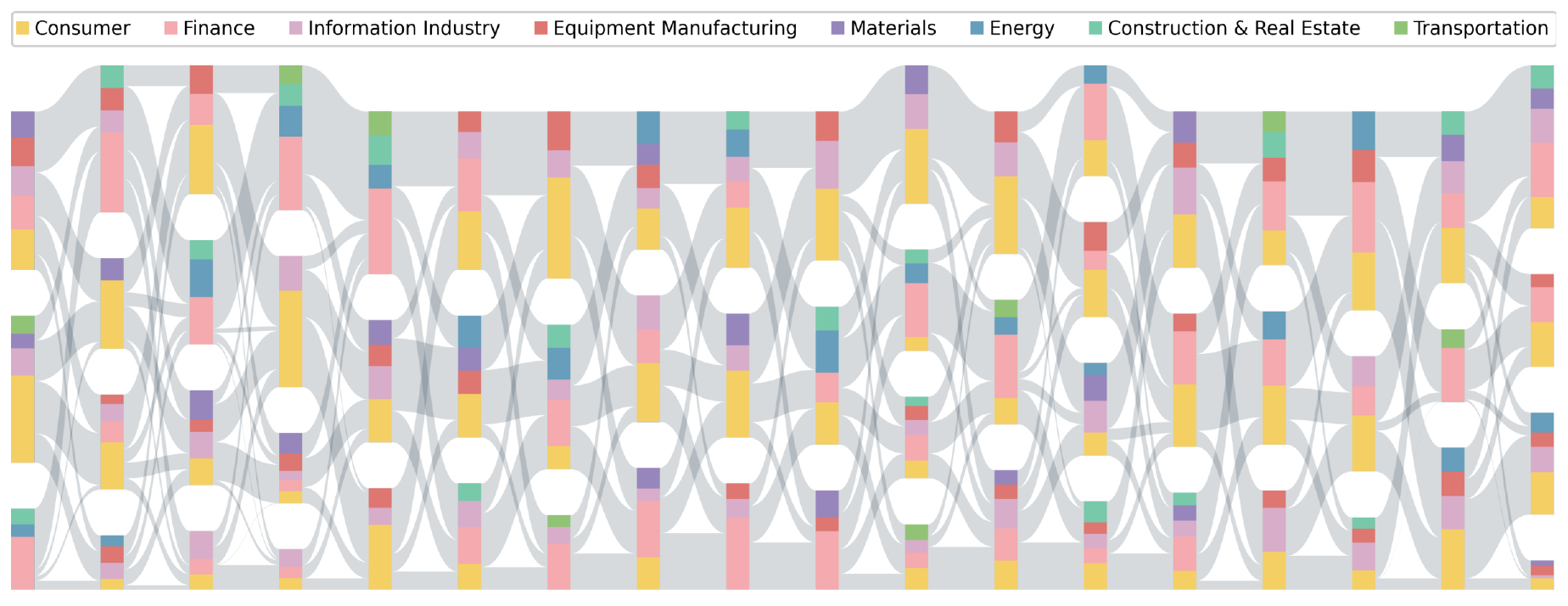}
\caption{{\bf Network community evolution with industrial elements across 18 consecutive time windows.} Communities are visualized as blocks of vertices and connected by curves to visualize transitions between communities. The flow of the curves indicates the dynamics of relationships between nodes and communities, while the width represents the number of nodes. The color of each community corresponds to the distribution of industries to which the nodes in the community belong.}
\label{fig:sankey}
\end{figure}

Furthermore, we analyzed the evolutionary dynamics of communities and industry distribution, as illustrated in Figure \ref{fig:sankey}. Each rectangle denotes a community, with the length of the rectangle indicating the size of the community and the color within the rectangle representing the industry distribution. To provide a more explicit depiction, we sort the proportion of stocks from each industry within each community in descending order. Subsequently, we iteratively selected industries until the cumulative number of stocks exceeded 70\%  of that community's total stocks. This approach retained the significant economic sectors while disregarding less dominant industries. 

The results from community partitioning and evolution reveal four distinct patterns: (1) The number of communities typically falls within the range of 3 to 4, indicating a relatively stable partitioning of the stock market at each time step, with significant interplay among the communities. (2) Stock movement between communities is frequent, suggesting that community influences might vary across different instances. (3) At the industry level, specific sectors aggregate within the same community. For example, the Finance sector tends to cluster together with Construction \& Real Estate. This alignment aligns with reality as the Real Estate industry carries financial attributes. Additionally, implicit industry relationships have been found, with Energy and Construction \& Real Estate also showing a tendency to aggregate. Conversely, other sectors, like Information Technology, often exhibit dispersion across different communities.

\section{Disscussion}
The construction of stock networks provides an effective tool for analyzing financial markets. This study proposes an autoencoder-based stock network inference method to extract the interconnections among stocks from daily price data. We explore its applications in stock price prediction, network topological analysis, evolutionary community detection, and portfolio selection.

Many recent studies have utilized stock networks for financial market analysis. Huang et al.\cite{huang2009network} constructed networks based on logarithmic stock returns, while Huang et al.\cite{huang2016financial} used Value at Risk to establish connections. However, these methods require manual determination of network thresholds or involve filtering approaches, making the construction of stock networks less flexible. In contrast, our approach employs SNIF  for encoding and inference, LSTM for time series feature extraction, and GCN for relation inference. This facilitates the automatic construction of networks from trading data, effectively revealing potential relationships among stocks. Such inference methodology offers a versatile framework applicable to stock networks and adaptable to constructing networks from other time series data, such as brain functional networks.

Moreover, we explore the effectiveness of the stock network in various applications. We present conclusions for four applications: in stock price prediction, our approach outperforms existing methods in prediction accuracy; in network topological analysis, we define movement dispersion and observe negative correlations between movement dispersion and degree, clustering coefficient, and degree assortativity, and a positive correlation with average path length; in evolutionary community detection, we uncover relationships between communities and industries, analyzing industry evolution patterns; in portfolio selection, we achieve the best results in selecting the top-k stocks based on centrality indicators. Besides these applications, our method-generated network can also be applied to constructing an index based on network structure and analyzing financial network stability.

Our study provides an auto-inferred stock network and validates its rationality through four applications. It demonstrates its effectiveness in price and return prediction, unearthing connections among stocks and aiding in a better understanding of financial market operating mechanisms and patterns.

\bibliographystyle{unsrt}

\bibliography{cas-refs}

\renewcommand{\thefigure}{A\arabic{figure}} 
\renewcommand{\thetable}{A\arabic{table}}
\setcounter{figure}{0}
\setcounter{table}{0} 

\section*{Appendix}

\newpage
\begin{table}[h]
\caption{{\bf The SNIF(LSTM) performance of temporal stock network movement prediction decoder.} The best results are shown in bold.}
\label{prediction table}
\centering
\begin{tblr}{
  cells = {c},
  cell{1}{2} = {c=2}{},
  cell{1}{4} = {c=2}{},
  cell{1}{6} = {c=2}{},
  hline{1-3,21-22} = {-}{},
}
     & ACC            &                & AUC            &                & Precision      &                \\
     & SNIF           & LSTM           & SNIF           & LSTM           & SNIF           & LSTM           \\
1    & \textbf{0.491} & 0.468          & \textbf{0.499} & 0.492          & \textbf{0.534} & 0.510          \\
2    & \textbf{0.547} & 0.516          & \textbf{0.558} & 0.516          & \textbf{0.623} & 0.472          \\
3    & \textbf{0.530} & 0.516          & \textbf{0.538} & 0.516          & \textbf{0.525} & 0.495          \\
4    & \textbf{0.521} & 0.512          & \textbf{0.520} & 0.512          & \textbf{0.520} & 0.501          \\
5    & \textbf{0.489} & 0.486          & \textbf{0.497} & 0.470          & \textbf{0.495} & 0.490          \\
6    & \textbf{0.562} & 0.555          & \textbf{0.579} & 0.570          & \textbf{0.553} & 0.539          \\
7    & \textbf{0.528} & 0.517          & \textbf{0.550} & 0.513          & \textbf{0.513} & 0.509          \\
8    & 0.505          & \textbf{0.507} & \textbf{0.557} & 0.530          & \textbf{0.571} & 0.551          \\
9    & 0.524          & \textbf{0.528} & 0.503          & \textbf{0.508} & 0.443          & \textbf{0.499} \\
10   & \textbf{0.550} & 0.510          & \textbf{0.565} & 0.520          & \textbf{0.596} & 0.500          \\
11   & \textbf{0.521} & 0.512          & \textbf{0.532} & 0.523          & \textbf{0.524} & 0.507          \\
12   & \textbf{0.532} & 0.508          & \textbf{0.526} & 0.489          & \textbf{0.478} & 0.443          \\
13   & \textbf{0.496} & 0.484          & \textbf{0.483} & 0.477          & \textbf{0.472} & 0.459          \\
14   & 0.494          & \textbf{0.506} & \textbf{0.491} & \textbf{0.491} & 0.430          & \textbf{0.437} \\
15   & \textbf{0.535} & 0.522          & \textbf{0.542} & 0.528          & \textbf{0.531} & 0.495          \\
16   & 0.541          & \textbf{0.556} & \textbf{0.565} & 0.524          & \textbf{0.477} & 0.475          \\
17   & 0.504          & \textbf{0.519} & \textbf{0.524} & 0.523          & 0.498          & \textbf{0.512} \\
18   & \textbf{0.562} & 0.532          & \textbf{0.532} & 0.522          & \textbf{0.489} & 0.452          \\
Mean & \textbf{0.524} & 0.514          & \textbf{0.531} & 0.512          & \textbf{0.515} & 0.491          
\end{tblr}
\end{table}

\begin{figure}[htbp]
\centering
\includegraphics[width=1 \linewidth]{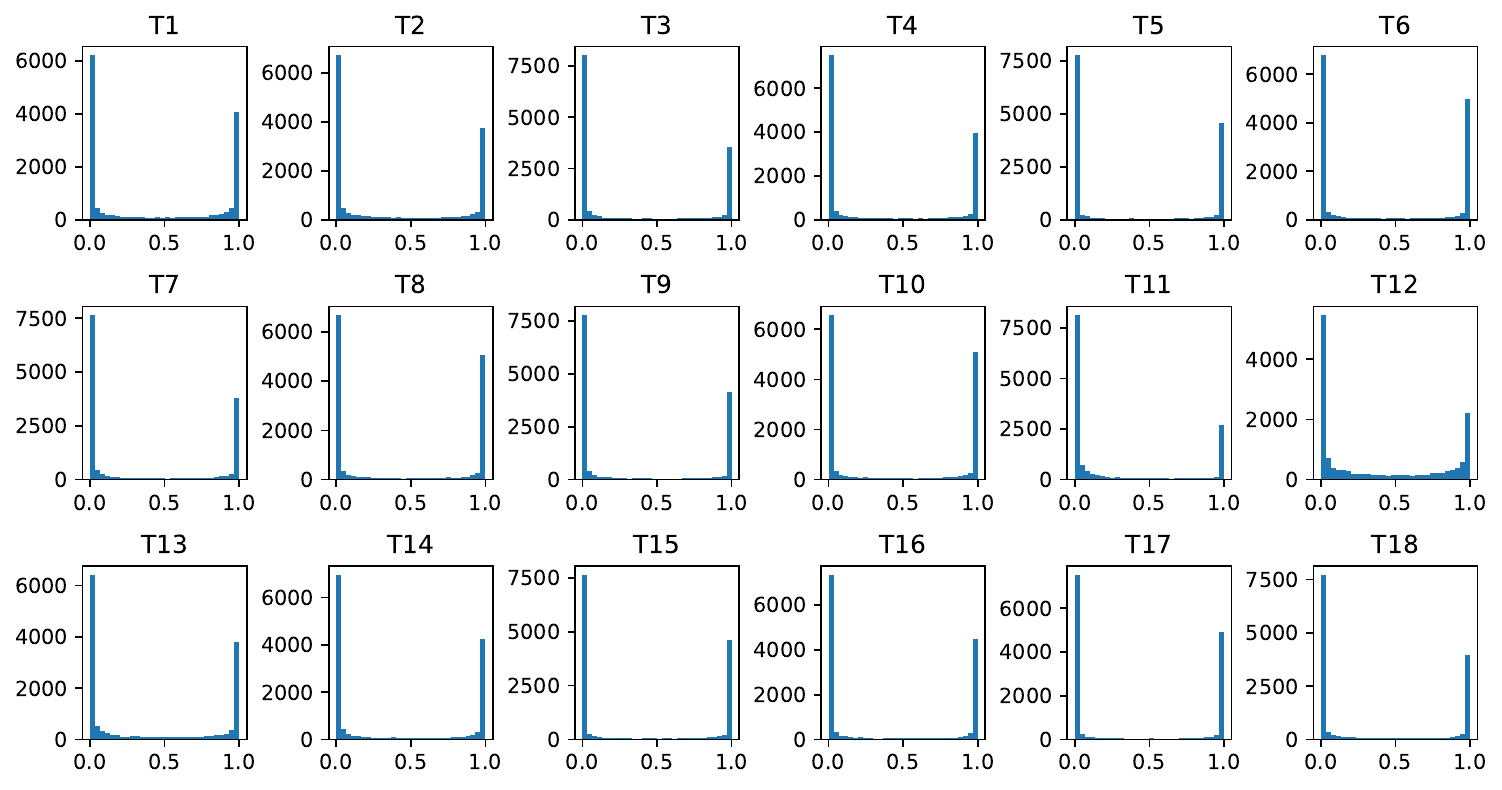}
\caption{{\bf The distribution of the edge weight.} For values greater than 0.5, we assume that there is an edge.}
\label{fig:edge_weight}
\end{figure}

\clearpage
\begin{figure}[h]
\centering
\includegraphics[width=0.8 \linewidth]{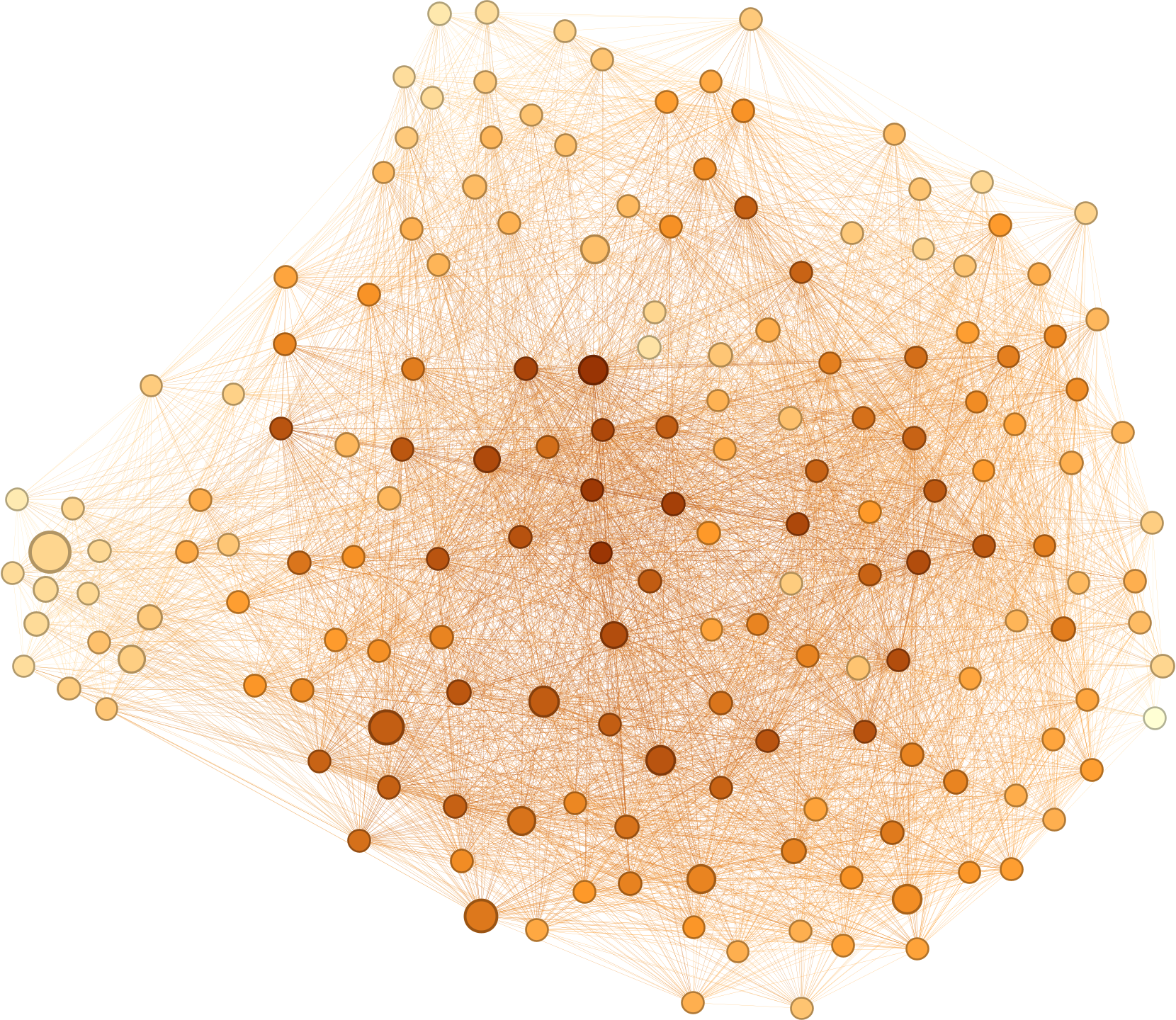}
\caption{{\bf The visualization of the stock network in the last time window.} The size of the node indicates the market capitalization of the stock, and the larger the node, the larger the corresponding market capitalization of the stock. The color of the node represents the degree, and the darker the color, the greater the degree of the node.}
\label{fig:stock17}
\end{figure}

\clearpage
\begin{table}[h]
\centering
\caption{{\bf The stock tickers and their corresponding industries.}}
\label{tab:stock_ticker}
\resizebox{0.95\linewidth}{10.5cm}{
\begin{tabular}{cccccc}
\hline
Stock Ticker & Industry & Stock Ticker & Industry & Stock Ticker & Industry \\
 \hline
600188 & Energy & 601088 & Energy & 601225 & Energy\\
 \hline
601898 & Energy & 600011 & Energy & 600025 & Energy\\
 \hline
600674 & Energy & 600795 & Energy & 600886 & Energy\\
 \hline
601985 & Energy & 600028 & Energy & 600346 & Energy\\
 \hline
601808 & Energy & 601857 & Energy & 002756 & Materials\\
 \hline
600019 & Materials & 002460 & Materials & 600111 & Materials\\
 \hline
600362 & Materials & 600547 & Materials & 603993 & Materials\\
 \hline
002493 & Materials & 002648 & Materials & 002709 & Materials\\
 \hline
600426 & Materials & 603260 & Materials & 603806 & Materials\\
 \hline
000786 & Materials & 002271 & Materials & 600585 & Materials\\
 \hline
000876 & Consumer Goods & 002311 & Consumer Goods & 002714 & Consumer Goods\\
 \hline
300498 & Consumer Goods & 600887 & Consumer Goods & 603288 & Consumer Goods\\
 \hline
000568 & Consumer Goods & 000596 & Consumer Goods & 000858 & Consumer Goods\\
 \hline
002304 & Consumer Goods & 600132 & Consumer Goods & 600519 & Consumer Goods\\
 \hline
600600 & Consumer Goods & 603369 & Consumer Goods & 002032 & Consumer Goods\\
 \hline
002050 & Consumer Goods & 000338 & Consumer Goods & 000625 & Consumer Goods\\
 \hline
002594 & Consumer Goods & 002920 & Consumer Goods & 600104 & Consumer Goods\\
 \hline
600660 & Consumer Goods & 600741 & Consumer Goods & 601238 & Consumer Goods\\
 \hline
601633 & Consumer Goods & 601689 & Consumer Goods & 601799 & Consumer Goods\\
 \hline
300347 & Consumer Goods & 300529 & Consumer Goods & 300595 & Consumer Goods\\
 \hline
600763 & Consumer Goods & 603882 & Consumer Goods & 603833 & Consumer Goods\\
 \hline
000963 & Consumer Goods & 002001 & Consumer Goods & 002007 & Consumer Goods\\
 \hline
002821 & Consumer Goods & 300122 & Consumer Goods & 300142 & Consumer Goods\\
 \hline
300601 & Consumer Goods & 600085 & Consumer Goods & 600196 & Consumer Goods\\
 \hline
600276 & Consumer Goods & 600436 & Consumer Goods & 002027 & Social Services\\
 \hline
603899 & Social Services & 000069 & Social Services & 601888 & Social Services\\
 \hline
000768 & Equipment Manufacturing & 600760 & Equipment Manufacturing & 601989 & Equipment Manufacturing\\
 \hline
601766 & Equipment Manufacturing & 300316 & Equipment Manufacturing & 300450 & Equipment Manufacturing\\
 \hline
300014 & Equipment Manufacturing & 300274 & Equipment Manufacturing & 600089 & Equipment Manufacturing\\
 \hline
600406 & Equipment Manufacturing & 600438 & Equipment Manufacturing & 600884 & Equipment Manufacturing\\
 \hline
601877 & Equipment Manufacturing & 000157 & Equipment Manufacturing & 600031 & Equipment Manufacturing\\
 \hline
601100 & Equipment Manufacturing & 002008 & Equipment Manufacturing & 002414 & Equipment Manufacturing\\
 \hline
600050 & Public Utilities & 600029 & Transportation & 600115 & Transportation\\
 \hline
601006 & Transportation & 601021 & Transportation & 601111 & Transportation\\
 \hline
601919 & Transportation & 002120 & Transportation & 600233 & Transportation\\
 \hline
600018 & Transportation & 000001 & Finance & 600000 & Finance\\
 \hline
600015 & Finance & 600016 & Finance & 600036 & Finance\\
 \hline
600926 & Finance & 601009 & Finance & 601166 & Finance\\
 \hline
601169 & Finance & 601229 & Finance & 601288 & Finance\\
 \hline
601328 & Finance & 601398 & Finance & 601818 & Finance\\
 \hline
601939 & Finance & 601988 & Finance & 601998 & Finance\\
 \hline
000166 & Finance & 000776 & Finance & 002736 & Finance\\
 \hline
300059 & Finance & 600061 & Finance & 600837 & Finance\\
 \hline
601211 & Finance & 601688 & Finance & 601788 & Finance\\
 \hline
601878 & Finance & 601881 & Finance & 601901 & Finance\\
 \hline
601318 & Finance & 601336 & Finance & 601601 & Finance\\
 \hline
601628 & Finance & 601117 & Construction \& Real Estate & 601186 & Construction \& Real Estate\\
 \hline
601618 & Construction \& Real Estate & 601668 & Construction \& Real Estate & 601669 & Construction \& Real Estate\\
 \hline
601800 & Construction \& Real Estate & 000002 & Construction \& Real Estate & 600048 & Construction \& Real Estate\\
 \hline
600383 & Construction \& Real Estate & 601155 & Construction \& Real Estate & 300628 & Information Industry\\
 \hline
002371 & Information Industry & 600584 & Information Industry & 000100 & Information Industry\\
 \hline
000725 & Information Industry & 000733 & Information Industry & 002179 & Information Industry\\
 \hline
002241 & Information Industry & 002475 & Information Industry & 002600 & Information Industry\\
 \hline
002841 & Information Industry & 300408 & Information Industry & 300433 & Information Industry\\
 \hline
600183 & Information Industry & 002230 & Information Industry & 002410 & Information Industry\\
 \hline
300033 & Information Industry & 300496 & Information Industry & 600570 & Information Industry\\
 \hline
600588 & Information Industry & 600845 & Information Industry & 002555 & Information Industry\\
\hline
\end{tabular}}
\end{table}

\end{document}